\PassOptionsToPackage{english}{babel}
\documentclass[reprint,aps,prd,twocolumn,showpacs,amssymb,amsfonts,superscriptaddress,longbibliography,nofootinbib]{revtex4-1}  
\usepackage[english]{babel}
\usepackage{graphicx}  
\usepackage{dcolumn}   
\usepackage{bm}        
\usepackage{bbm}        
\usepackage{amssymb}   
\usepackage{amsmath}
\usepackage{mathtools}
\usepackage[utf8]{inputenc}
\usepackage[squaren]{SIunits}
 
\usepackage{changes}

\definecolor{myurlcolor}{rgb}{0,0,0.7}
\definecolor{myrefcolor}{rgb}{0.8,0,0}
\usepackage{hyperref}
\hypersetup{colorlinks, linkcolor=myrefcolor,
citecolor=myurlcolor, urlcolor=myurlcolor}

\usepackage{cleveref}
\crefrangeformat{equation}{#3(#1 #4--#5 #2)#6}

\usepackage{comment}
\usepackage{soul} 

\newcommand{\nn}{\nonumber}

\renewcommand\({\left(}
\renewcommand\){\right)}
\renewcommand\[{\left[}
\renewcommand\]{\right]}

\newcommand{\ra}{\rightarrow}

\def\lsim{\raise 0.4ex\hbox{$<$}\kern -0.8em\lower 0.62
ex\hbox{$\sim$}}

\def\gsim{\raise 0.4ex\hbox{$>$}\kern -0.7em\lower 0.62
ex\hbox{$\sim$}}

\def\lbar{{\hbox{$\lambda$}\kern -0.7em\raise 0.6ex
\hbox{$-$}}}

\newcommand\eq[1]{Eq.~(\ref{#1})}
\newcommand\eqs[2]{Eqs.~(\ref{#1}) and (\ref{#2})}

\newcommand\pa{\partial}
\newcommand\p{\partial}

\newcommand\ee{\end{equation}}
\newcommand\be{\begin{equation}}
\def\bea{\begin{array}}
\def\eea{\end{array}}\def\ea{\end{array}}
\newcommand\ees{\end{eqnarray}}
\newcommand\bees{\begin{eqnarray}}
\def\nn{\nonumber}

\def\f{\phi}
\def\D{\Delta}

\def\d{\delta}

\def\eps{\epsilon}

\def\dslash{\hspace{-1mm}\not{\hbox{\kern-2pt $\partial$}}}
\def\Dslash{\not{\hbox{\kern-2pt $D$}}}
\def\pslash{\not{\hbox{\kern-2.1pt $p$}}}
\def\kslash{\not{\hbox{\kern-2.3pt $k$}}}
\def\qslash{\not{\hbox{\kern-2.3pt $q$}}}

\newcommand{\vp}{{\bf p}}

\def\p1{{\bf p}_1}
\def\p2{{\bf p}_2}
\def\k1{{\bf k}_1}
\def\k2{{\bf k}_2}

\newcommand{\eMN}{\eta^{\mu\nu}}

\newcommand{\dddM}{\kern 0.2em \raise 1.9ex\hbox{$...$}\kern -1.0em \hbox{$M$}}
\newcommand{\dddQ}{\kern 0.2em \raise 1.9ex\hbox{$...$}\kern -1.0em \hbox{$Q$}}
\newcommand{\dddI}{\kern 0.2em \raise 1.9ex\hbox{$...$}\kern -1.0em\hbox{$I$}}
\newcommand{\dddJ}{\kern 0.2em \raise 1.9ex\hbox{$...$}\kern-1.0em
\hbox{$J$}}
\newcommand{\dddcalJ}{\kern 0.2em \raise 1.9ex\hbox{$...$}\kern-1.0em
\hbox{${\cal J}$}}

\newcommand{\dddO}{\kern 0.2em \raise 1.9ex\hbox{$...$}\kern -1.0em
\hbox{${\cal O}$}}
\def\dddz{\raise 1.5ex\hbox{$...$}\kern -0.8em \hbox{$z$}}
\def\dddd{\raise 1.8ex\hbox{$...$}\kern -0.8em \hbox{$d$}}
\def\dddbd{\raise 1.8ex\hbox{$...$}\kern -0.8em \hbox{${\bf d}$}}
\def\ddbd{\raise 1.8ex\hbox{$..$}\kern -0.8em \hbox{${\bf d}$}}
\def\dddx{\raise 1.6ex\hbox{$...$}\kern -0.8em \hbox{$x$}}

\newcommand{\mpl}{M_{\rm P}}

\begin{document}
\selectlanguage{english}

\title{Revisiting the algebraic structure of the generalized uncertainty principle}

\author{Matteo Fadel}
\email{fadelm@phys.ethz.ch}
\affiliation{Department of Physics, ETH Z\"{u}rich, 8093 Z\"{u}rich, Switzerland}

\author{Michele Maggiore}
\email{michele.maggiore@unige.ch}
\affiliation{D\'epartement de Physique Th\'eorique and Center for Astroparticle Physics,\\
Universit\'e de Gen\`eve, 24 quai Ansermet, CH--1211 Gen\`eve 4, Switzerland}


\date{\today}

\begin{abstract}
We compare different formulations of the  generalized uncertainty principle that have an underlying algebraic structure. We show that the formulation by Kempf, Mangano and Mann (KMM) [\href{http://dx.doi.org/10.1103/PhysRevD.52.1108}{Phys. Rev. D \textbf{52} (1995)}], quite popular for phenomenological studies, satisfies the Jacobi identities only for spin zero particles. In contrast, the formulation proposed earlier by one of us (MM) [\href{http://dx.doi.org/10.1016/0370-2693(93)90785-G}{Phys. Lett. B \textbf{319} (1993)}] has an underlying algebraic structure valid for particles of all spins, and is in this sense more fundamental. The latter is also much more constrained, resulting into only two possible solutions, one expressing the existence of a minimum length, and the other expressing a form of quantum-to-classical transition. We also discuss how this more stringent algebraic formulation has an intriguing physical interpretation in terms of a discretized time at the Planck scale.
\end{abstract}

\maketitle

\section{Introduction}

The idea of a generalized uncertainty principle (GUP), of the form
\be\label{GUP1}
\Delta x\, \gsim\, \frac{\hbar}{2\Delta p} +{\rm const.}\times  G_N\Delta p
\ee
(where  $G_N$ is Newton's constant), and its associated minimum resolvable length, emerges  from the computation of scattering amplitudes in string theory 
at Planckian energies~\cite{Veneziano:1986zf,Amati:1987wq,Amati:1988tn,Gross:1987kza,Gross:1987ar}
and from Gedanken experiment with black holes and Hawking radiation~\cite{Maggiore:1993rv,Adler:1999bu} (although the idea of  a minimum length in gravity has a very long history, see~\cite{Hossenfelder:2012jw,Hagar:2014} for  historical discussions).
In these frameworks, the term proportional to  $G_N$  corresponds to a first-order quantum gravity correction, valid in the limit $\Delta p/(\mpl c)\ll 1$, where $\mpl = \sqrt{\hbar c / G_N}$ is the Planck mass. Therefore, \eq{GUP1} should be understood as 
\be\label{GUP2}
\Delta x\ge \frac{\hbar}{2\Delta p}\[ 1  +\beta_0 \(\frac{\Delta p}{\mpl c}\)^2 
+{\cal O}\(\frac{\Delta p}{\mpl c}\)^4\]\; ,
\ee
where    $\beta_0$ is  a dimensionless  constant [related to the constant  appearing in \eq{GUP1} by ${\rm const.}= \beta_0/(2c^3)$]. 

It is natural to ask whether there is an algebraic structure underlying the GUP, much as the canonical commutator $[x_i,p_j]=i\hbar\delta_{ij}$ underlies the standard Heisenberg uncertainty relation, and, in the affirmative case, if, under some reasonable assumptions, the algebraic structure is sufficiently constraining
to determine (almost) uniquely the structure of the higher-order terms in 
\eq{GUP1}. This question was first posed, and answered affirmatively, by one of us (MM) in \cite{Maggiore:1993kv}
(see also \cite{Maggiore:1993zu}). A  different answer to the same question was later provided by Kempf, Mangano and Mann (KMM) in \cite{Kempf:1994su}. 

In this paper we further elaborate on the algebraic formulations of the GUP. In section~\ref{sect:comp} we compare the MM and KMM formulations of the GUP, and argue that the former is more fundamental, as the resulting commutators satisfy the Jacobi identities in full generality, while, in the approach of KMM, it is implicitly assumed that the $x_i$ are the coordinates of a spin-zero particle. If the  existence of a minimum length should emerge from  a fundamental theory of quantum gravity as a basic property of space-time, it must hold independently of the type of particle used to probe it, and, in this sense, the formulation in \cite{Maggiore:1993kv} seems  more suitable  to emerge from a fundamental theory. As we will see, this formulation  is also more restrictive, fixing uniquely (modulo a sign)  the $[x_i,x_j]$ and the $[x_i,p_j]$ commutators, while the KMM approach involves an arbitrary function of momentum. Out of the two solutions allowed by the Jacobi identities in the approach of \cite{Maggiore:1993kv}, only one describes the existence of a minimum length. The other, which differs in a crucial sign,  was already mentioned in \cite{Maggiore:1993kv}, but received little attention.
In section~\ref{sect:quantumtoclass} we will discuss the latter solution in more detail, and we show that it can be seen as describing a transition from quantum to classical mechanics, with all commutators vanishing at a critical energy. In section~\ref{sect:kPoincare}, elaborating on results presented in \cite{Maggiore:2002qr}, we will see how the two solutions allowed by the Jacobi identities can be understood as emerging in a setting in which time becomes discrete at the Planck scale. In secttion~\ref{sect:composite} we will examine the effect of GUPs to composite objects, confirming previous results obtained in the KMM framework~\cite{Amelino-Camelia:2013fxa}, that indicate a strong suppression of GUP effects at macroscopic scales when the deformed commutators are applied to the constituent particles. Some further generalization of the algebraic structure underlying the GUP are discussed in Section~\ref{sect:basis}. Finally, section~\ref{sect:concl} summarizes our conclusions.

\section{Comparison of different algebraic approaches to the GUP}\label{sect:comp}

Let us begin by recalling the MM approach followed in Ref.~\cite{Maggiore:1993kv} to find an algebraic structure underlying the GUP. One starts by assuming that:
(1) the three-dimensional rotation group is not deformed, so the rotation generators ${\bf J}$  satisfy the undeformed commutation
relations $[J_i,J_j]=i\epsilon_{ijk}J_k$,
and coordinates and momenta satisfy the undeformed
commutation relations of spatial vectors, $\left[ J_i,x_j\right] =i\epsilon_{ijk}x_k,
\left[ J_i,p_j\right] =i\epsilon_{ijk}p_k$. (2) The momenta commutes among themselves: $\left[ p_i,p_j\right]
=0$, so that also the translation group is not deformed.
(3) The $\left[ x,x\right]$ and $\left[ x,p\right]$ commutators
depend on a deformation
parameter $\kappa$ with dimensions of mass. In the limit
$\kappa\ra\infty$ (that is, $\kappa c^2$ much larger than any energy in the problem), the
canonical commutation relations are recovered.
With these assumptions,
one is led to look for an expression for the  $[x,x]$ and $[x,p]$  commutators of the form
\begin{eqnarray}
\left[ x_i ,x_j \right] &= & \(\frac{\hbar}{\kappa c}\)^2 \,  a(p)\,
i\epsilon_{ijk}J_k\; , \label{xxa}
\\
\left[ x_i , p_j \right]   &= & i\hbar\,\delta_{ij}f(p)\; . \label{xpf}
\end{eqnarray}
Having assumed rotational invariance, the functions $a(p)$ and $f(p)$ (which are real and dimensionless) can depend on momentum only through it modulus $p\equiv |\vp|$; equivalently, they can be written as functions of 
$E/(\kappa c^2)$, where $E$ is defined by $E^2= p^2c^2+m^2c^4$.\footnote{We take $E$ as a notation for $(p^2c^2+m^2c^4)^{1/2}$. The actual dispersion relation between energy and momentum, in the context of the GUP, is also often modified, see e.g. \cite{Maggiore:1993zu}. In that case, we denote by ${\cal E}$ the actual energy of the system, whose relation to $E$ will be non-trivial, see sect.~\ref{sect:kPoincare}.\label{note:Enotation}} Compared to \cite{Maggiore:1993kv}, we keep $c$ explicit, rather than setting $c=1$, and we prefer to use $p$ instead of $E$ as the argument of the functions, to make more clear the relation with the KMM result in~\cite{Kempf:1994su}.
The angular momentum ${\bf J}$ is defined as  dimensionless, i.e. is in units of $\hbar$, while $\kappa$ has dimensions of mass and, eventually, will be identified with the Planck mass times a numerical constant. In principle, one could also add a term proportional to $p_ip_j$ to the right-hand side of \eq{xpf}. We will discuss in sect.~\ref{sect:basis} how this term can be eliminated. 
 
 We will work in the context of non-relativistic quantum mechanics. While there has been much work toward Lorentz-covariant deformed commutation relations (see~\cite{Hossenfelder:2012jw} for review) it is not obvious that this is the correct way to proceed. In fact, already in the undeformed case, the relativistic generalization is obtained in a different way through quantum field theory, rather than promoting $[x_i,p_j]=i\hbar\delta_{ij}$ to something like $[x^{\mu},p^{\nu}]=i\hbar\eMN$.

The functions $a(p)$ and $f(p)$ can be constrained by imposing that the deformed commutators satisfy the  Jacobi identities. The non-trivial ones are 
$\left[ x_i,\left[ x_j,x_k\right]\right] +{\rm cyclic}=0$ and 
$\left[ x_i,\left[ x_j,p_k\right]\right]+ {\rm cyclic} =0$, and give the conditions~\cite{Maggiore:1993kv}
\bees
\frac{d a(p)}{dp}\, {\bf p}\cdot{\bf J}&=&0\; ,\label{dadp}\\
\frac{f(p)}{p}\frac{d f(p)}{dp}&=&-\frac{a(p)}{\kappa^2 c^2}\label{dfdp}\; .
\ees
The crucial point, that is at the basis of the difference between the results of MM, Ref.~\cite{Maggiore:1993kv}, and KMM, Ref.~\cite{Kempf:1994su}, is the following. If we restrict to  orbital angular momentum so that ${\bf J}={\bf L}$, then ${\bf p}\cdot{\bf J}=0$ automatically, and \eq{dadp} is  satisfied without the need of imposing $da/dp=0$. Therefore, one remains with just one relation between $a(p)$ and $f(p)$, given by \eq{dfdp}. One can for instance choose 
$f(p)$ arbitrarily, and then $a(p)$ follows. This is the approach implicitly taken by KMM, where $f(p)$ is eventually arbitrarily chosen to have the form $f(p)=1+\beta p^2$ \cite{Kempf:1994su}.  However, if we want to interpret the GUP as a fundamental property of quantum gravity, its validity should not be restricted to spin-zero particles, but should hold generally. For a generic spin, ${\bf p}\cdot{\bf J}$ is non-vanishing (it is indeed the helicity of the particle). Therefore, in Ref.~\cite{Maggiore:1993kv}, it was rather imposed that $da/dp=0$, so $a(p)$ must be a constant. With a rescaling of $\kappa$, we can then set $a(p)=\pm 1$. Consider first the solution $a(p)=-1$ (we will come back to the other solution
 in sect.~\ref{sect:quantumtoclass}). Then \eq{dfdp} integrates to
\be\label{fpalpha}
f(p)=\sqrt{\alpha+\frac{p^2}{\kappa^2 c^2}}\; ,
\ee
where $\alpha$ is an integration constant. This constant can be fixed requiring that, at energies $E\ll\kappa c^2$, or momenta $p\ll\kappa c$, the standard uncertainty principle is recovered [this also fixes the plus sign in front of the square root in \eq{fpalpha}]. If we work using momentum as variable, as in \eq{fpalpha}, the natural choice is then $\alpha=1$, so that the Heisenberg uncertainty relation is recovered at $p=0$, and 
\eqs{xxa}{xpf} become
\begin{eqnarray}
\left[ x_i ,x_j \right] &= & -\(\frac{\hbar}{\kappa c}\)^2 \,
i\epsilon_{ijk}J_k\; , \label{gupxx}
\\
\left[ x_i , p_j \right]   &= & i\hbar\,\delta_{ij}\, \sqrt{1+\frac{p^2}{\kappa^2 c^2}}\; . \label{gupxp}
\end{eqnarray}
Actually, in Ref.~\cite{Maggiore:1993kv} was made a different choice for the integration constant, which resulted in
\be
f(p)=\sqrt{1+\frac{p^2+m^2c^2}{\kappa^2 c^2}}\; ,
\ee
i.e., in terms of $E$,
\be
f(E)=\sqrt{1+\frac{E^2}{\kappa^2 c^4}}\; ,
\ee
so that the $[x,x]$ commutator is still given by \eq{gupxx}, while
\be \label{gupxpE}
\left[ x_i , p_j \right]   =  i\hbar\,\delta_{ij}\, \sqrt{1+\frac{E^2}{\kappa^2 c^4}}\; .
\ee
This choice was made because a GUP of this form emerges naturally in the context of the $\kappa$-deformed Poincar\'e algebra~\cite{Maggiore:1993zu}, as we will see in  section~\ref{sect:kPoincare}. Both \eqs{gupxp}{gupxpE} are logically possible, within this framework. At $p\ll\kappa c$, \eq{gupxpE} reduces to
\be
\left[ x_i , p_j \right]   \simeq   i\hbar\,\delta_{ij}\, \sqrt{1+\frac{m^2}{\kappa^2}}\; ,
\ee
corresponding to a mass-dependent rescaling of $\hbar$, while \eq{gupxp}  reduces to
$[x_i,p_i]\simeq i\hbar \delta_{ij}$. If $m$ is interpreted as the mass of an elementary particle, and $\kappa$ is of the order of the Planck mass, $m/\kappa$ is  negligibly small, and \eqs{gupxp}{gupxpE} are basically the same. We will examine in section~\ref{sect:composite} the situation for macroscopic objects, whose mass can easily exceed the Planck mass.

We can now compare MM's \eqs{gupxx}{gupxp}, or \eqs{gupxx}{gupxpE}, with the deformed algebra proposed by KMM, which reads\footnote{See their Eqs.~(75) and (77); we denote by $F(p^2)$ the function that they call $f(p^2)$, to avoid a notation conflict with the function  that enters in \eq{xpf} and that, following the notation in \cite{Maggiore:1993kv}, we denote as  $f(p)$.} ~\cite{Kempf:1994su}
\bees
\left[ x_i ,x_j \right] &= &  -2\hbar \frac{dF(p^2)}{d(p^2)} i(x_i p_j-x_jp_i)\; ,\label{kmm1}\\
\left[ x_i , p_j \right] &=&  i\hbar\,\delta_{ij} [1+F(p^2)]\; .\label{kmm2}
\ees
Furthermore, the  orbital angular momentum (in units of $\hbar$), in the approach of Ref.~\cite{Kempf:1994su} is expressed in terms of coordinates and momenta as
\be
L_{ij}=\frac{1}{\hbar }\, \frac{1}{1+F(p^2)}\, (x_i p_j-x_jp_i)\; ,
\ee
so that \eqs{kmm1}{kmm2} can be rewritten as
\bees
\left[ x_i ,x_j \right] &= &  -2\hbar^2 [1+F(p^2)] \frac{dF(p^2)}{dp^2} i\eps_{ijk}L_k\; ,\label{kmm3}\\
\left[ x_i , p_j \right] &=&  i\hbar\,\delta_{ij} [1+F(p^2)]\; .\label{kmm4}
\ees
Comparison with \eqs{xxa}{xpf} shows, first of all, that Ref.~\cite{Kempf:1994su} is implicitly assuming $J_i$ to be the same as the orbital angular momentum $L_i$. Therefore, the constraint $da/dp=0$ is lost.
Writing $1+F(p^2)=f(p)$, we then get back \eqs{xxa}{xpf}, with $a(p)$ expressed in terms of $f(p)$ through  the remaining constraint (\ref{dfdp}).

Eventually, Ref.~\cite{Kempf:1994su} specializes to the simple choice $F(p^2)=\beta p^2$, with $\beta$ a constant [written
in terms of a dimensionless constant $\beta_0$ as $\beta=\beta_0/(\mpl^2 c^2)$], and this form of the GUP has become very popular for phenomenological studies. It should be stressed, however, that, apart from its simplicity, there is no real justification for this choice. The strength of the MM approach of Ref.~\cite{Maggiore:1993kv} is that it requires that the Jacobi identities hold for particles with all spins, and not only for spin-zero particles, and this fixes uniquely the function $a(p)$ and $f(p)$ (modulo a sign, that we will discuss further below, and the different possible choices of integration constant $\alpha$ in $f(p)$, as we have discussed). In contrast, in the approach of KMM, the algebra is only valid for spin-zero particles and, as a result, one loses a constraint and the function $f(p)$ becomes completely arbitrary. Any function of the form $f(p)=1 +\beta p^2 +{\cal O}(p^4)$ would reproduce
\eq{GUP2}, and the particular truncation $f(p)=1 +\beta p^2$ has no special motivation. In particular, there is no reason why such an expression should hold as $p$ approaches the Planck scale.\footnote{Many subsequent works, inspired by the approach in \cite{Kempf:1994su}, have proposed variants of the GUP by choosing other specific forms of the function $f(p)$ with supposedly desirable properties, see e.g.~\cite{Nouicer:2007jg,Pedram:2011gw,Chung:2019raj,Petruzziello:2020een}. However,
all these approaches suffer from the same basic arbitrariness of the proposed functional form.}

\begin{figure}
    \centering
    \includegraphics[width=\columnwidth]{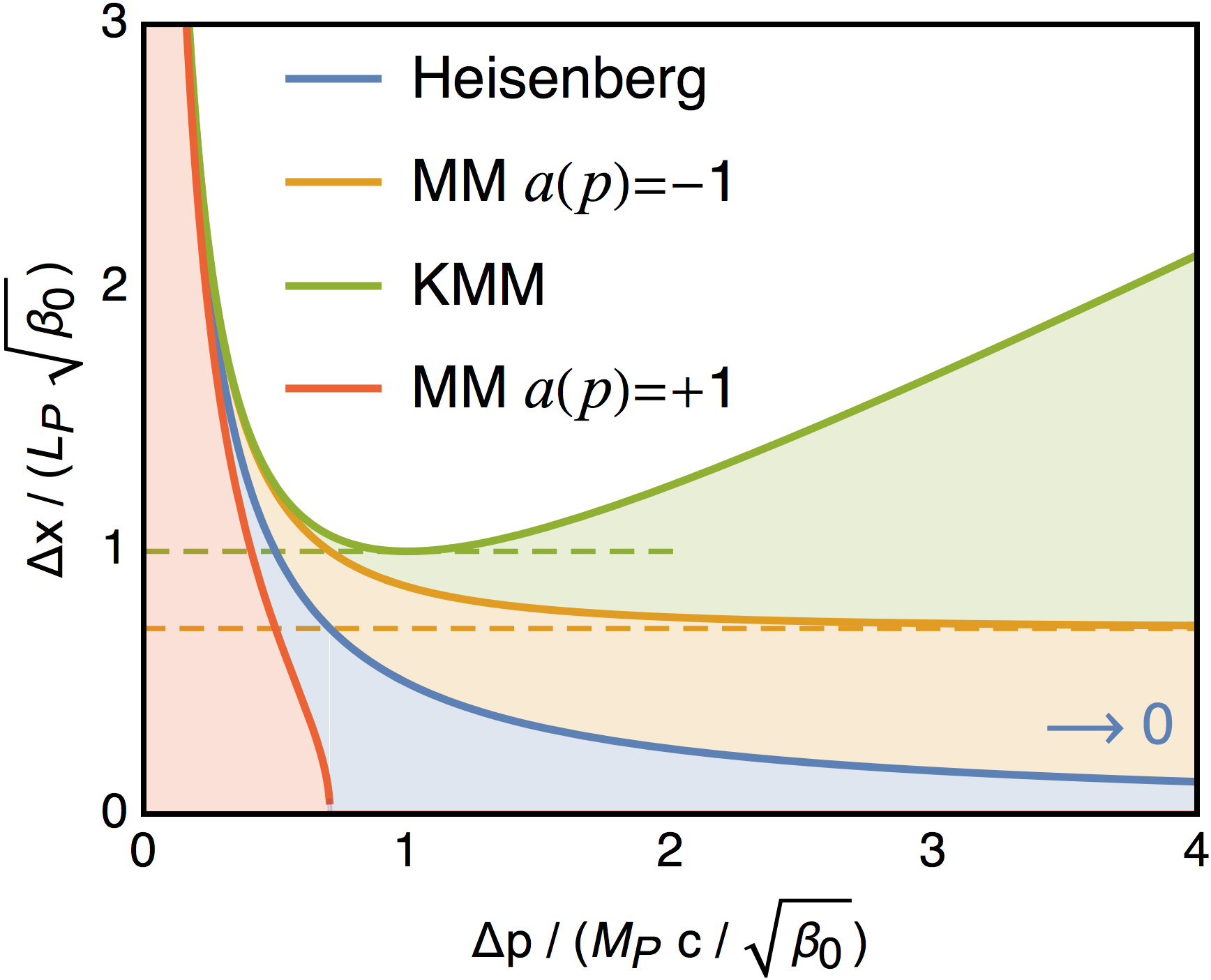}
    \caption{\textbf{Bound on $\Delta x$ as a function of $\Delta p$ for different uncertainty principles.} Blue: the Heisenberg uncertainty relation allows for arbitrarily small $\Delta x$ as $\Delta p$ increases. Orange: the MM GUP Eq.~\eqref{finalDxDpmu0} obtained from Eq.~\eqref{GUPsqrtbeta} predicts a minimum $\Delta x$ for large $\Delta p$. Green: the KMM GUP Eq.~\eqref{urkmm} obtained from Eq.~\eqref{comKMM} predicts a minimum $\Delta x$ for a finite $\Delta p$, after which the position uncertainty increases again. Red: the other MM GUP obtained from Eq.~\eqref{gupxpEminus}, which can be associated to a quantum-to-classical transition where $[x_i,p_j]=0$ when $\Delta p$ is larger than a critical value. Shaded regions are regions excluded by the corresponding uncertainty relation.}
    \label{fig:compGUP}
\end{figure}

Observe  that, away from the lowest order in $\Delta p/M_p c$, in which all these algebraic structures reproduce \eq{GUP2}, the MM GUP obtained from \eq{gupxp}, and the KMM obtained from \eq{kmm4} with $F(p^2)=\beta p^2$, are completely different.
To make contact with the notation commonly used in the the context of \eq{kmm4}, we introduce also in \eq{gupxp} a constant  $\beta_0$ from
\be
\frac{1}{\kappa^2}=\frac{2\beta_0}{\mpl^2}\; ,
\ee
and $\beta=\beta_0/(\mpl^2 c^2)$, so \eq{gupxp} can be written as
\bees
\left[ x_i , p_j \right]   &=&  i\hbar\,\delta_{ij}\, \sqrt{1+2\beta_0\frac{p^2}{\mpl^2 c^2}}\nn\\
 &=&  i\hbar\,\delta_{ij}\, \sqrt{1+2\beta p^2}\; .\label{GUPsqrtbeta}
\ees
The uncertainty principle derived from this commutation relation is
\be
\Delta x_i\Delta p_j\, \geq\, \frac{\hbar}{2}\delta_{ij}\, 
\langle \,\sqrt{ 1+2\beta p^2 }\,  \rangle\; ,
\ee
where $\langle\ldots\rangle$ denotes the quantum expectation value on a given state.
We can write this more explicitly expanding the square root as
\be \label{expasqrt}
\sqrt{ 1+2\beta p^2 } =\sum_{n=0}^{\infty} c_n (2\beta p^2)^n\; ,
\ee
where we used the generalized binomial coefficient
\be
c_n=  {1/2 \choose n} = 
\frac{ (-1)^n (2n)! }{ 2^{2n} (1-2n) (n!)^2} \;.
\ee
Then
\bees
\Delta x_i\Delta p_j  &\, \geq\, & \frac{\hbar}{2}\delta_{ij}\,  \sum_{n=0}^{\infty} c_n (2\beta)^n 
\langle p^{2n}  \rangle\nn\\
&\, \geq\, &  \frac{\hbar}{2}\delta_{ij}\,  \sum_{n=0}^{\infty} c_n (2\beta)^n \langle p^{2}  \rangle^n\nn\\
&= &  \frac{\hbar}{2}\delta_{ij}\,  \sum_{n=0}^{\infty} c_n (2\beta)^n [ \langle p\rangle^2+ (\Delta p)^2 ]^n\nn\\
 &= &  \frac{\hbar}{2} \delta_{ij}\, 
 \sqrt{1+2\beta [ \langle p\rangle^2+(\Delta p)^2] }\; ,
\ees
so that \eq{gupxp} gives (see Fig.~\ref{fig:compGUP}, yellow line)
\bees
\Delta x_i\Delta p_j &\, \geq\,& \frac{\hbar}{2}\delta_{ij}\, \sqrt{1+2\beta_0 \frac{ \langle p\rangle^2+(\Delta p)^2 }{\mpl^2 c^2} }\nn\\
&\, \geq\,& \frac{\hbar}{2}\delta_{ij}\, \sqrt{1+2\beta_0 \frac{(\Delta p)^2 }{\mpl^2 c^2} }
\; . \label{finalDxDpmu0}
\ees
For values of  $\Delta p$ much smaller than $\mpl c$,
this becomes 
\be
\Delta x_i\Delta p_j\, \gsim\, \frac{\hbar}{2}\delta_{ij}
\( 1+ \beta_0\frac{(\Delta p)^2}{\mpl^2 c^2} \)\; ,
\ee
which is of the form (\ref{GUP1}).  In contrast, in the opposite limit
$\Delta p\gg\mpl c$, \eq{finalDxDpmu0} gives
\be
\Delta x_i\Delta p_j\, \gsim\, \hbar \delta_{ij} \(\frac{\beta_0}{2}\)^{1/2}\, \, \frac{\Delta p}{\mpl c} \;,
\ee
which implies 
\begin{equation}\label{minDxA}
\Delta x \, \gsim  \, L_p \(\frac{\beta_0}{2}\)^{1/2} \; ,
\end{equation}
where $L_P = \sqrt{\hbar G / c^3}$ is the Planck length.
Therefore, the minimum uncertainty for $\Delta x$ saturates to a value of order of $L_P$ times a numerical factor that depends on $\beta_0$. 

The same computation performed for \eq{gupxpE} gives
\bees\label{finalDxDpmu0E}
\Delta x_i\Delta p_j  &\, \geq\, &\frac{\hbar}{2}\delta_{ij}\, \sqrt{1+2\beta_0 \frac{ E^2+ (\Delta p)^2c^2}{\mpl^2 c^4} }\nn\\
&\, \geq\, &\frac{\hbar}{2}\delta_{ij}\, \sqrt{1+2\beta_0 \frac{m^2c^2+ (\Delta p)^2}{\mpl^2 c^2} } \; ,
\ees
and in the limit $\Delta p\gg \mpl c$ (with $m<\mpl$) we recover Eq.~\eqref{minDxA}.

In contrast, the KMM GUP obtained from Eq.~\eqref{kmm4} with $F(p)=\beta p^2$ has a different behaviour. To see this, note that the commutator
\be
\left[ x_i , p_j \right]   =  i\hbar\,\delta_{ij}\, \[ 1+\beta_0 \(\frac{p}{\mpl c}\)^2 \]  \label{comKMM}
\ee
results in the GUP (see Fig.~\ref{fig:compGUP}, green line)
\be \label{urkmm}
\Delta x_i\Delta p_j\,\geq \,\frac{\hbar}{2}\delta_{ij}
 \[ 1+\beta_0 \frac{ (\Delta p)^2 }{ (\mpl c)^2 } \] \;.
\ee
Similarly to Eq.~\eqref{finalDxDpmu0}, this expression also implies the existence of a minimal position uncertainty 
\begin{equation}
    \Delta x \geq L_P \sqrt{\beta_0} \;,
\end{equation}
which is attained for $\Delta p = M_p c / \sqrt{\beta_0}$. However, for $\Delta p\gg\mpl c$ we have
\be
\Delta x\, \gsim\, \frac{\hbar}{2}\, \frac{\beta_0}{\mpl c}\, \frac{\Delta p}{\mpl c}\; ,
\ee
meaning that the position uncertainty now grows without bounds as the momentum uncertainty increases.

\section{Generalized uncertainty principle  and quantum-to-classical transition}\label{sect:quantumtoclass}

We now turn the attention to the solution of \eqs{dadp}{dfdp} obtained by setting $a(p)=+1$. In this case,
\eq{fpalpha} becomes 
\be\label{fpalphaminus}
f(p)=\sqrt{\alpha-\frac{p^2}{\kappa^2 c^2}}\; .
\ee
With the choice  of integration constant $\alpha=1$,  we have
\begin{eqnarray}
\left[ x_i ,x_j \right] &= & \(\frac{\hbar}{\kappa c}\)^2 \,
i\epsilon_{ijk}J_k\; , \label{gupxxminus}
\\
\left[ x_i , p_j \right]   &= & i\hbar\,\delta_{ij}\, \sqrt{1-\frac{p^2}{\kappa^2 c^2}}\; . \label{gupxpminus}
\end{eqnarray}
Similarly, with the choice of integration constant analogous to that leading to \eq{gupxpE},  we get
\be \label{gupxpEminus}
\left[ x_i , p_j \right]   =  i\hbar\,\delta_{ij}\, \sqrt{1-\frac{E^2}{\kappa^2 c^4}}\; .
\ee
This solution is  intriguing. Since the square root on the right-hand side is a decreasing function of  momentum (always smaller than one), it no longer produces a term in the uncertainty principle that induces  a minimum observable length, which was the initial motivation of these investigations. Rather, it describes a situation in which a system becomes  more and more classical as its momentum (or its energy) approaches a critical value, of the order of the Planck scale, where eventually the commutator vanishes. Beyond this energy, one can then simply set 
$[x_i,p_j]=0$, by continuity. Note that, if  set $f(p)=0$ for $p$ larger than the critical value $\kappa c$ [or for $E$ larger than the critical value $\kappa c^2$, if we use \eq{gupxpEminus}], \eq{dfdp} also implies $a(p)=0$ for $p>\kappa c$ (or, respectively, $E>\kappa c^2$), so the geometry becomes commutative.
It therefore describes a sort of quantum-to-classical transition, in which, beyond a critical energy or momentum, the $[x,p]$ commutator vanishes (see Fig.~\ref{fig:compGUP}, red line), and also $[x_i,x_j]=0$.\footnote{This solution was already mentioned in \cite{Maggiore:1993kv}, where it was observed that it leads to a vanishing $[x,p]$ commutator at a limiting energy scale. The possibility that deformed commutation relations can result in a vanishing commutator at some energy scale was  rediscovered much later in \cite{Magueijo:2002am,Jizba:2009qf}, and, recently, in Ref.~\cite{Petruzziello:2020een}, which indeed proposes, on purely phenomenological grounds, the commutator \eq{gupxpminus}, that was originally found in \cite{Maggiore:1993kv} from algebraic considerations.
}

For $E$ smaller than $\kappa c^2$, but close to it, \eq{gupxpEminus} becomes
\be \label{gupxpEminusnearEc}
\left[ x_i , p_j \right]   \simeq  i\hbar\,\delta_{ij}\, \sqrt{2}\, \(1-\frac{E}{\kappa c^2}\)^{1/2}\; .
\ee
The energy dependence on the right-hand side  has the typical form of the behavior of an order parameter at a second order phase transition, with a critical index equal to $1/2$. The prototype example  of this is the  magnetization as a function of the temperature in  the Ising model, which, in the mean field approximation, is given by
\be\label{MTIsing}
M(T) \simeq M_0\( 1-\frac{T}{T_c}\)^{1/2}\; .
\ee
Note also that, contrary to some interpretation in the literature, an equation such as  \eq{gupxpminus} or \eq{gupxpEminus} does not necessarily imply the existence of a maximum momentum, since the commutator can just be set to zero by continuity above the critical value. This is similar to the fact  that \eq{MTIsing} does not imply the existence of a maximum temperature for the Ising model; but simply that $M(T)=0$ for $T>T_c$.

\section{Algebraic GUP and time discretization at the Planck scale}\label{sect:kPoincare}

The algebraic formulation of the GUP given in \eqs{gupxx}{gupxpE}, as well as that in \eqs{gupxxminus}{gupxpEminus}, has an interesting relation with deformations of the 
Poincar\'e algebra, and to a discretization of time at the Planck scale~\cite{Maggiore:1993zu}. We review this idea here, following the discussion presented in
Ref.~\cite{Maggiore:2002qr}.

\subsection{Discretized spatial dimensions and deformed Poincar\'e symmetry}\label{sect:discretespace}

Let us begin by considering a 1+1 dimensional system with a continuous time variable, while space is discretized on a regular lattice, with lattice spacing $d$.
A wave equation, such as a Klein-Gordon equation, becomes in this framework (setting, for simplicity, $\hbar=c=1$ in the intermediate steps)
\be\label{KG1}
\( -\frac{1}{c^2}\pa_t^2 +\D_x^2 -m^2 \)\f =0 \;,
\ee
where 
\be\label{discretedx}
\D_x\f (t,x) =\frac{\f (t, x+d)-\f (t,x-d)}{2d}
\ee
is a discretization of the spatial derivative.
The dispersion relation that follows is\footnote{As mentioned in footnote~\ref{note:Enotation},
we keep $E$ as a notation for $(p^2c^2+m^2c^4)^{1/2}$ and now use the symbol ${\cal E}$ for the actual energy of the system.}
\be\label{disp2}
{\cal E}^2=\frac{\sin^2 (d\, p)}{d^2} +m^2\; . 
\ee
Momentum is periodically identified, and  there is a maximum energy that can be carried
by a wave solution, 
${\cal E}_{\rm max}=(d^{-2}+m^2)^{1/2}$.
If we  replace the speed of light $c$ (that here we have set to unity) by a speed $v<1$, this
equation  describes the propagation of phonons in $1+1$ dimensions. 

A system described by \eq{KG1} has an underlying Lie algebra with two symmetry generators:
the generator $H$ of continuous time translations and the
generator $P$ of discrete spatial translations, satisfying the Lie algebra $[H,P]=0$, supplemented by the
identification $P\sim P+2\pi/d$. The symmetry under boosts is broken, and no
generator is associated to it.
However, it was observed long ago \cite{Bonechi:1992qf} that this system has an alternative description in terms of  a deformed algebra: one introduces
also the boost generator $K$, and considers the algebra
\bees
\[P,H\]&&=0\, ,\hspace{10mm} \[K,P\]=iH\, ,\label{PH1}  \\
&&\[K,H\]=i\,\frac{\sin (2dP)}{2d}\; .\label{PH2}
\ees
In the limit $d\ra 0$ this reduces to the Poincar\'e algebra of a 1+1
continuous relativistic system. 
The algebra given by \eqs{PH1}{PH2}, however, is 
well defined also at finite $d$, since  the commutators  obey the Jacobi identities.  Eqs.~(\ref{PH1}, \ref{PH2}) provide an example of a  deformed algebra, in which  $d$ is the deformation parameter. Its relevance, in connection with a system described by \eq{KG1}, is that
this deformed algebra has a quadratic Casimir $C_2$ given by
\be
C_2 =H^2-\frac{\sin^2 (d P)}{d^2}\; ,
\ee
Therefore, the  dispersion
relation (\ref{disp2})  is simply the condition
$C_2=m^2$, and in this sense this deformed Poincar\'e algebra can be
considered as the symmetry of a relativistic system living in discrete
one-dimensional space and
continuous time.\footnote{Note that the particular choice (\ref{discretedx}) for the discretization of the spatial derivative is irrelevant here. A different discretization would just produce a different function of $P$ on the right-hand side of \eq{PH2}, but 
the Jacobi identities are  satisfied even if, on the   right-hand side of \eq{PH2}, we have  an arbitrary function of $P$.}

Comparing the Lie algebra and the deformed algebra descriptions of the symmetries in
this system we see that, in the Lie algebra approach, when $d\neq 0$, there are only two
generators, $H$ and $P$; $d=0$ is a point of enhanced
symmetry, where a new symmetry transformation,  boosts, emerges, and the corresponding  generator $K$ suddenly appears. In the deformed algebra description, instead, we always have a description in terms  of
three generators $H,P,K$ even for finite
$d$, but we pay this with a non-linear structure. The value  $d=0$ is a special point at which  this algebraic structure linearizes. 

\subsection{Discreteness of time and the algebraic GUP}

Consider next a system in which  time is discrete, in steps of size $\tau$, and space is
continuous. We begin with $1+1$ dimensions, and we consider
the equation (still setting for the time being $\hbar=c=1$)
\be 
(-\D_t^2+\pa_x^2-m^2)\f =0\; ,
\ee
where 
\be\label{discretedx}
\D_t\f (t,x) =\frac{\f (t+\tau, x)-\f (t-\tau,x)}{2\tau}\; ,
\ee
and $\tau$ is the discrete time step.
The corresponding dispersion relation is
\be\label{disp0}
\frac{\sin^2 \tau{\cal E}}{\tau^2}= p^2+m^2\; .
\ee
We can find a quantum algebra description of this system, just by exchanging the role of $H$ and $P$ in \eqs{PH1}{PH2}, and replacing the spatial step  $d$ with the temporal step $\tau$, 
\bees
\[P,H\]&&=0\, ,\hspace{10mm} \[K,H\]=iP\; ,\label{HP1}  \\
&&\[K,P\]=i\,\frac{\sin (2\tau H)}{2\tau}\; .\label{HP2}
\ees
The  quadratic  Casimir operator of this algebra is
\be
C_2 =\frac{\sin^2 (\tau H)}{\tau^2}-P^2\; .
\ee
so the dispersion relation (\ref{disp0}) corresponds again to the condition $C_2=m^2$, with ${\cal E}$ identified with the eigenvalue of the operator $P^0$ and $p$ with the eigenvalue of the operator $P$.

The above construction can be generalized to the case of  discrete time and more continuous spatial dimensions. Indeed,  \eqs{HP1}{HP2} are just the restriction to $1+1$ dimensions of the $\kappa$-deformed Poincar\'e algebra, 
a deformation of
the Poincar\'e algebra that can be written in any number of
dimensions as it follows~\cite{Lukierski:1991,Lukierski:1992dt}. All commutators involving the angular momentum $J_{ij}$
are the same as in the undeformed
Poincar\'e algebra. Hence, the group of spatial
rotations is not deformed. Similarly, for space-time
translations still holds 
$[P_{\mu},P_{\nu}]=0$. The commutators involving the boosts
$K_i=J_{i0}$  are instead
\be
\[K_i,H\]=iP_i\, ,\hspace{5mm} [K_i,P_j]=i\d_{ij}
\frac{\sin (2\tau H)}{2\tau}\; ,\label{lnr1}
\ee
and
\bees
\[K_i,K_j\]&=&-iJ_{ij}\cos (2\tau H)\nn\\
&&-
i\tau^2P^k(P_iJ_{jk}+P_jJ_{ki}+P_kJ_{ij})\; .\label{lnr2}
\ees
The quadratic Casimir is
\be
C_2 =\frac{\sin^2 (\tau H)}{\tau^2}-{\bf P}^2\;  ,
\ee
and therefore $C_2=m^2$ gives
\be\label{dispvp}
\frac{\sin^2 \tau{\cal E}}{\tau^2}= \vp^2+m^2\; .
\ee
The relation between the $\kappa$-deformed Poincar\'e algebra and the algebraic formulation of the GUP emerges in the following way~\cite{Maggiore:1993zu,Maggiore:2002qr}.
In standard quantum mechanics,  we
quantize a particle imposing 
\be\label{UP}
[x_i,p_j]=i \d_{ij}\; .
\ee 
(Recall that we are temporarily  setting $\hbar=1$. We will restore $\hbar$ and $c$ at the end). In momentum space the operator $x_i$ can  then be
represented as 
\be\label{xiUP}
x_i=i\, \frac{\pa}{\pa p_i}\; ,
\ee
and  the velocity of the
particle in the Heisenberg representation is  given by
\be
\dot{x}_i=i\, [H,x_i]=\frac{\pa {\cal E}}{\pa p_i}\; .
\ee
For the case of a discretized spatial dimension, discussed in section~\ref{sect:discretespace}, using \eq{disp2} 
(and setting  for simplicity $m=0$) gives
\be
\frac{\pa {\cal E}}{\pa p_i} = \hat{p}_i\cos (ap)\; ,
\ee
where $\hat{p}_i$ is the unit vector in the direction of $\vp$. This
 is  the standard result for the group velocity of a massless particle 
on a regular spatial lattice. 

However, the same procedure applied to the case of a discrete time dimension gives immediately a puzzling  result.  If we assume the validity of \eq{UP}, and therefore of \eq{xiUP}, 
using \eq{dispvp} we get
\be\label{3.5}
v_i=\frac{\pa {\cal E}}{\pa p_i} = \(\frac{p_i}{\sqrt{p^2+m^2}}\)\, \frac{1}{\cos
(\tau{\cal E})}\; .
\ee
The term in parenthesis is just the standard expression for the
velocity in terms of momentum. However, the cosine at the denominator 
makes no sense, and if we  take \eq{3.5} as an expression for the
velocity,  we  find that $v>1$, and even diverges when $\tau{\cal E}$
approaches $\pi/2$.

Clearly, \eq{xiUP} cannot be the correct expression for the position operator when time is discrete and the dispersion relation has the form (\ref{dispvp}). Rather, a natural approach is to define the
position operator requiring that the relation between 
velocity and momentum, $v_i=p_i/\sqrt{p^2+m^2}$, stays unchanged, 
so that, in particular, $v$ increases monotonically from zero to the speed of light (that we have set here to one) as momentum ranges from zero to infinity. This can be obtained defining, in momentum space,
\be\label{3.6}
x_i=i \cos(\tau{\cal E})\frac{\pa}{\pa p_i}\; .
\ee
By construction we now have
\be
v_i=i[H,x_i]=\cos (\tau{\cal E}) \frac{\pa {\cal E}}{\pa p_i}
=\frac{p_i}{\sqrt{p^2+m^2}}\; .
\ee
Then, 
using \eq{3.6}, we can compute explicitly the $[x_i,x_j]$ and 
$[x_i,p_j]$ commutators, and we find (restoring the correct powers of $\hbar$ and $c$)
\bees
\[ x_i, x_j\] 
& = & i(c\tau)^2J_{ij}\, ,\label{xxm}\\
\[ x_i, p_j\] & = & i \hbar
\delta_{ij}\, \sqrt{1-\frac{\tau^2}{\hbar^2}({\bf p}^2c^2+m^2c^4)}\; ,\label{xpm}
\ees
where we have defined 
\be\label{J}
J_{ij}=-i\( p_i\frac{\pa}{\pa p_j}-p_j\frac{\pa}{\pa p_i}\)\; .
\ee
We thus recovered the modified algebra given in \eqs{gupxxminus}{gupxpEminus}, with the identification $\tau=\hbar/(\kappa c^2)$.

This result is quite remarkable. It shows that the GUP given by \eqs{gupxxminus}{gupxpEminus} [which, apart from a choice of integration constant, is one of the only two possible expressions fixed by algebraic arguments, once one correctly takes into account that
\eq{dadp} requires $da/dp=0$ in order for the argument to be valid for particles with arbitrary spin] has an intriguing physical interpretation as the natural form of the uncertainty principle when the  time variable  becomes discrete at the Planck scale.  Therefore, a discretization of time at the Planck scale implies that a system becomes classical as its energy  approaches the Planck energy, since there the commutator $[x_i,p_j]$ vanishes.

We can now ask what setting corresponds to the other allowed form of the algebraic uncertainty principle, given by \eqs{gupxx}{gupxpE}. As discussed in Ref.~\cite{Maggiore:2002qr}, this GUP emerges automatically when we start from  an euclidean KG equation (here for simplicity in one spatial dimension)
\be
(\D_t^2+\pa_x^2-m^2)\f =0\; .
\ee
The corresponding dispersion relation is 
\be
-\frac{\sin^2 \tau{\cal E}}{\tau^2}= p^2+m^2\; .
\ee
We then perform a Wick rotation back  to Minkowski space, which is equivalent to   
transforming ${\cal E}\ra -i{\cal E}$, and 
the dispersion relation becomes
\be\label{disp}
\frac{\sinh^2 \tau E}{\tau^2}= p^2+m^2\; .
\ee
Formally \eqs{disp0}{disp} are  related by $\tau\ra i\tau$.
Substituting $\tau\ra i\tau$ into \eqs{HP1}{HP2} we therefore
find a deformation of the Poincar\'e algebra whose
Casimir reproduces the dispersion relation (\ref{disp}),
\bees
\[P,H\]&&=0\, ,\hspace{10mm} \[K,H\]=iP\; ,\label{euP1}  \\
&&\[K,P\]=i\,\frac{\sinh (2\tau H)}{2\tau}\; .\label{euP2}
\ees
This algebra can be generalized to more spatial dimensions, and it is the other version of the $\kappa$-Poincar\'e algebra found in~\cite{Lukierski:1992dt,Lukierski:1993wxa}. The deformed commutators are
\be\label{elnr1}
[K_i,H]=iP_i\, ,\hspace{5mm} [K_i,P_j]=i\d_{ij}
\frac{\sinh (2\tau H)}{2\tau}\; ,
\ee
\bees
[K_i,K_j]&=&-iJ_{ij}\cosh (2\tau H)\nn\\
&&+
i\tau^2P^k(P_iJ_{jk}+P_jJ_{ki}+P_kJ_{ij})\; ,\label{elnr2}
\ees
with all other commutators undeformed, and the quadratic Casimir
\be
C_2 =\frac{\sinh^2 (\tau H)}{\tau^2}-{\bf P}^2\; .
\ee
In this case, proceeding as above, one realizes that the correct definition of the position operator 
is~\cite{Maggiore:1993zu,Maggiore:2002qr}\footnote{In general, the definition of the position operator includes also a term proportional to $p_i$, that ensures the hermiticity with respect to the scalar product invariant under the Poincar\'e group (or its deformations). For the undeformed Poincar\'e group, this gives $x_i=i[\pa/\pa_i-(p_i/2p_0^2)]$. The corresponding expressions for the deformed Poincar\'e group 
can be found in \cite{Maggiore:1993zu}. In any case, as long as $[p_i,p_j]=0$, as we assume here, the term proportional to $p_i$ does not affect the $[x_i,x_j]$ and the $[x_i,p_j]$ commutators, and, for simplicity, we omit them.}
\be
x_i=i\cosh (\tau E)\frac{\pa}{\pa p_i}
\ee
and, computing the $[x_i,x_j]$ and $[x_i,p_j]$ commutators, one finds the GUP in the form given in \eqs{gupxx}{gupxpE}.

\section{GUP for composite systems}\label{sect:composite}

A well-know problem concerning the GUP is how to extend it from microscopic degrees of freedom to composite  objects and, eventually, to  macroscopic objects. The non-linearity of the GUP commutators implies that the commutator obeyed by the center of mass position and total momentum of a composite object is not the same as that of the fundamental constituents (see below). A first question, then, is to what `fundamental' constituents the GUP, in any of the forms discussed above,  is supposed to apply. If the GUP emerges from a fundamental theory of gravity, it is natural to assume that it will apply to the elementary excitations of that theory. The next issue is what happens to composite systems and, in particular, to macroscopic objects.
In the realm of elementary particles, terms such as $E^2/(\kappa^2c^4)$ in eqs.~(\ref{gupxpE}) or
(\ref{gupxpEminus}) represent small corrections, if we take $\kappa$ of order of $\mpl$, and $E\ll \mpl c^2$. However, for macroscopic objects, these terms are huge. The Planck mass correspond to about $10^{-5}\, {\rm gr}$, so for a macroscopic object of mass $m$ even $m/\kappa$ can be large. Naively,  this seems to imply that GUP effects become very large for macroscopic objects. This, however,  is not necessarily the case, as it was shown in \cite{Amelino-Camelia:2013fxa} (see also the Supplementary material in \cite{Pikovski:2011zk}) for the KMM form of the GUP. The argument is as follows. Consider a macroscopic body made of $N$ particles, taken for simplicity to have all the same mass, and labeled by an index $\alpha=1,\ldots, N$, with coordinates $x^i_{\alpha}$ and momenta $p^i_{\alpha}$.
The center of mass coordinates $X^i$ and the total momentum $P^i$ of the composite system are defined as
\be
X^i=\frac{1}{N}\sum_{\alpha=1}^N x^i_{\alpha}\, ,\qquad
P^i=\sum_{\alpha=1}^N p^i_{\alpha}\, .
\ee
Assuming a GUP of the form (\ref{kmm4}) with
$F(p^2)=\beta p^2$, we have
\be\label{xapb}
\left[ x^i_{\alpha} , p^j_{\alpha'} \right] =  i\hbar \delta^{ij}\,\delta_{\alpha\alpha'} (1+\beta p_{\alpha}^2)\, .
\ee
Then
\bees
\[ X^i,P^j\] &=& \frac{1}{N}\sum_{\alpha=1}^N\sum_{\alpha'=1}^N\, \[x^i_{\alpha}, p^j_{\alpha'}\]\nn\\
&=&i\hbar \delta^{ij} \frac{1}{N}\sum_{\alpha=1}^N\,(1+\beta p_{\alpha}^2)\nn\\
&=&i\hbar \delta^{ij}\[ 1+ \frac{\beta}{N}\sum_{\alpha,\beta=1}^N p_{\alpha}^2\]\, .\label{GAC1}
\ees
For an object moving in an quasi-rigid manner, in order of magnitude  $p_{\alpha}\sim P/N$, where $P$ is the modulus of the total momentum, and therefore
\be
\[ X^i,P^j\] =i\hbar \delta^{ij}\[ 1+ {\cal O}( \frac{\beta  P^2 }{N^2})\]\, .
\ee
Therefore, for a macroscopic object, $\beta$ in \eq{xapb} has been replaced by $\beta/N^2$ and, for $N$ of the order of the number of constituent of a macroscopic object, this quantity is utterly negligible. More precisely, 
\eq{GAC1} can be rewritten as  \cite{Amelino-Camelia:2013fxa}
\be
\[ X^i,P^j\]=i\hbar \delta^{ij}\[ 1+ \frac{\beta  P^2 }{N^2} +
 \frac{\beta}{N}\sum_{\alpha,\beta=1}^N \(p_{\alpha}^2-\frac{P^2}{N^2}\)
 \]\, .
 \ee
 The last term is the variance of $p_{\alpha}^2$, multiplied by $\beta/N$, and vanishes for an exactly rigid body. For non-rigid objects, depending on the details of the internal structure, one could hope that the variance leaves, overall, an effect that scales as $(\beta P^2/N^{\alpha})$, with $\alpha$ a power in the range $0<\alpha<2$~\cite{Kumar:2019bnd}. In any case, the effect of the GUP appears to be strongly suppressed for macroscopic objects.
 
Essentially the same derivation can be adapted to the GUP in the forms  (\ref{gupxp}) or (\ref{gupxpminus}) [or in the forms written in terms of energy, \eqs{gupxpE}{gupxpEminus}]. Using for definiteness  \eq{GUPsqrtbeta}, expanding the square root as in \eq{expasqrt} and replacing $p_{\alpha}=P/N$ for an exactly rigid object, we get 
\bees
\[ X^i,P^j\] &=&i\hbar \delta^{ij} \frac{1}{N}\sum_{\alpha=1}^N\,\sqrt{1+2\beta p_{\alpha}^2}\nn\\
&=&i\hbar \delta^{ij}\, \frac{1}{N}\sum_{\alpha=1}^N\,\sum_{n=0}^{\infty} c_n (2\beta p_{\alpha}^2)^n\nn\\
&=&i\hbar \delta^{ij}\, \frac{1}{N}\sum_{\alpha=1}^N\,\sum_{n=0}^{\infty} c_n (2\beta P^2/N^2)^n\nn\\
&=&i\hbar \delta^{ij}\, \sqrt{1+\frac{2\beta P^2}{N^2}}\; .
\ees
In summary, if the GUP applies to the constituents particles of a rigid macroscopic body, the same form of the modified commutator applies to the center of mass position and momentum with a rescaling $\beta \rightarrow \beta/N^2$.

\section{Further extensions of the algebraic structure}\label{sect:basis}

When discussing the consequences of a GUP, one should not forget that the structure of the commutators is only one side of the aspect, and the dynamics, e.g. the form of the Hamiltonian, is another. These two are tied together by the possibility of redefining variables. Consider, for instance, a one-dimensional system,  characterized by a commutator of the form
\be\label{xpfofp}
[x,p]=i\hbar f(p)\, .
\ee
In momentum space, the operator $x$ can be represented as 
\be
x\ra i\hbar f(p)\frac{\pa}{\pa p}\, .
\ee
Then, defining  $k(p)$ from requiring
\be
f(p)\frac{dk}{dp}=1\, ,
\ee
at the corresponding operator level we get the canonical commutator
\be
[x,k]=i\hbar f(p)\frac{\pa}{\pa p} k=i\hbar\, . \label{comXK}
\ee 

From the point of view of the dynamics, consider the Hamiltonian of the system to be
\be
H[x,p]=\frac{p^2}{2m}+V(x) \;.
\ee 
In terms of  the $(x,p)$ variables we have a `normal' Hamiltonian, but a deformed commutator (\ref{xpfofp}). On the other hand, if we rather use the variables 
$(x,k)$ we have a canonical commutation relation (\ref{comXK}), at a price of a modified Hamiltonian
\be
H[x,k]=\frac{p^2(k)}{2m}+V(x)  \;.
\ee
For a general function $f(p)$, $H[x,k]$ contains arbitrarily high powers of $k$ and therefore, in coordinate space, contains spatial derivatives of arbitrarily high order [i.e., it is non-local, in the field-theoretical sense]. 

Note however that, if we take as Hamiltonian 
\be
H=\frac{k^2(p)}{2m}+V(x) 
\ee 
in terms of the $(x,p)$ variables, this would look as a very complicated quantum system described by deformed commutation relations and nonlocal Hamiltonian; even if, once one passes to the $(x,k)$ variables, one realizes that this is  just a normal quantum system, with a standard local  Hamiltonian and canonical commutation relation. This trivial example shows that the algebraic structure of the commutation relations is only one side of the coin, and must always be examined together with the dynamics of the system (see also the discussion in \cite{Pedram:2011aa,Hossenfelder:2012jw}).

The above example, where the commutator could be reduced to the underformed one, is however specific to a one-dimensional system. In more dimensions, in fact, the situation is more complex. First of all, one also has to consider the $[x_i,x_j]$ commutator, which is non-trivial. Furthermore, there is also the possibility of adding a term $p_ip_j$ to the right-hand side of \eq{xxa}. The most general algebra consistent with rotational invariance is thus
\begin{eqnarray}
\left[ x_i ,x_j \right] &= & \(\frac{\hbar}{\kappa c}\)^2 \,  a(p)\,
i\epsilon_{ijk}J_k\; , \label{xxapipj}
\\
\left[ x_i , p_j \right]   &= & i\hbar\,\[ f(p)\delta_{ij} +g(p)p_ip_j\] \; , \label{xpfpipj}
\end{eqnarray}
which depends on three functions $a(p)$, $f(p)$ and $g(p)$.
In the momentum representation, the position operator can then be written as
\be
x_i\ra i\hbar \[ f(p) \frac{\pa}{\pa p_i} + g(p) p_ip_k \frac{\pa}{\pa p_i} \]\, .
\ee
The non-trivial Jacobi identities now reduce to two differential equations for the three functions
$a(p)$, $f(p)$ and $g(p)$, and these
are no longer enough to find a unique  solution (modulo a sign); rather, after imposing them, one remains with a solution parametrized by a free function. 

However, consider the transformation to a new momentum variable of the form $k_i=u(p)p_i$, which is the most general transformation consistent with rotational invariance. 
Then, for the associated operators we get
\bees\label{commxikj}
[x_i,k_j]&=& f(p)u(p)\delta_{ij} \nn\\
&&\hspace*{-10mm}+\[ \frac{u'(p)}{p} \( f(p)+g(p) p^2\) +g(p)u(p)\] p_ip_j\; .
\ees
We can then choose $u(p)$  so that the term proportional to $p_ip_j$ in \eq{commxikj} vanishes and, after renaming $f(p)u(p)$ as $f(p)$, we get back the structure in \eq{xpf}, with now $k_i$ as momentum variable. In this sense, one can always restrict to \eqs{xxa}{xpf}, possibly at the price of nonlocal  terms in the Hamiltonian, induced by the elimination of the term proportional to $p_ip_j$.

Alternatively, we can  restrict the freedom by imposing $a(p)=0$. In this case we get $[x_i,x_j]=0$, 
$[p_i,p_j]=0$, and then the Jacobi identities  fix the  $[x_i,p_j]$ 
commutator to~\cite{Ali:2009zq,Ali:2011fa}
\be
[x_i,p_j]=i
\hbar\, \[\delta_{ij}(1-\alpha p+\alpha^2 p^2) +p_ip_j\(-\frac{\alpha}{p}+3\alpha^2\)\]\; .
\ee

\section{Conclusions}\label{sect:concl}

In this paper we have revisited some aspects of the algebraic approach to the generalized uncertainty principle and to deformations of the commutation relations. We have pointed out an important difference between the approaches by MM~\cite{Maggiore:1993kv} and by KMM~\cite{Kempf:1994su}, emphasizing that the KMM approach implicitly selects a branch of solution of the Jacobi identities that only holds for spin-zero particles, and hence cannot be representative of fundamental properties of space-time. In contrast, requiring, as in \cite{Maggiore:1993kv}, that the algebraic formulation is consistent (in the sense of obeying the Jacobi identities) independently of the particle spin, provides a more fundamental approach, which also has some remarkable consequences. First, the algebraic structure becomes fixed, modulo a sign; in contrast to the KMM approach, where the $[x,p]$ commutator is an arbitrary function of momentum, which is then fixed on rather arbitrary grounds. Second, the two solutions that emerge (in correspondence with the two possible signs) describe quite different physics; one gives a GUP with a minimum length uncertainty of the type found in string theory or with Gedanken experiment with black holes; the other describes a system that, near the Planck scale, becomes classical. Moreover, these two solutions have a very intriguing interpretation, namely as the natural commutation relations obtained in theories where time becomes discrete at the Planck scale. The two different solutions emerge, in this context, either discretizing directly Minkowski time, or discretizing time in the Euclidean formulation, and then rotating back to Minkowski space.

Finally, we have also re-examined, in our context, the issue of GUP for composite system, confirming the conclusion that the GUP does not extend trivially to macroscopic objects. If the deformed commutator is applied at the level of the constituent particles, then its effects are suppressed by powers of the number of constituents, thus suggesting that the effects of GUP will be limited to the realm of elementary particles with Planckian energies.

\vspace{5mm}
\textbf{Acknowledgments. }
We thank Igor Pikovski and Yiwen Chu for  useful discussions. MF was supported by The Branco Weiss Fellowship -- Society in Science, administered by the ETH Z\"{u}rich.
The work  of MM is supported by the  Swiss National Science Foundation and  by the SwissMap National Center for Competence in Research.

\bibliographystyle{utphys}
\bibliography{refsGUP}

\providecommand{\href}[2]{#2}\begingroup\raggedright\begin{thebibliography}{10}

\bibitem{Veneziano:1986zf}
G.~Veneziano, ``{A Stringy Nature Needs Just Two Constants},''
  \href{http://dx.doi.org/10.1209/0295-5075/2/3/006}{{\em Europhys. Lett.}
  {\bfseries 2} (1986) 199}.

\bibitem{Amati:1987wq}
D.~Amati, M.~Ciafaloni, and G.~Veneziano, ``{Superstring Collisions at
  Planckian Energies},''
  \href{http://dx.doi.org/10.1016/0370-2693(87)90346-7}{{\em Phys. Lett. B}
  {\bfseries 197} (1987) 81}.

\bibitem{Amati:1988tn}
D.~Amati, M.~Ciafaloni, and G.~Veneziano, ``{Can Space-Time Be Probed Below the
  String Size?},'' \href{http://dx.doi.org/10.1016/0370-2693(89)91366-X}{{\em
  Phys. Lett. B} {\bfseries 216} (1989) 41--47}.

\bibitem{Gross:1987kza}
D.~J. Gross and P.~F. Mende, ``{The High-Energy Behavior of String Scattering
  Amplitudes},'' \href{http://dx.doi.org/10.1016/0370-2693(87)90355-8}{{\em
  Phys. Lett. B} {\bfseries 197} (1987) 129--134}.

\bibitem{Gross:1987ar}
D.~J. Gross and P.~F. Mende, ``{String Theory Beyond the Planck Scale},''
  \href{http://dx.doi.org/10.1016/0550-3213(88)90390-2}{{\em Nucl. Phys. B}
  {\bfseries 303} (1988) 407--454}.

\bibitem{Maggiore:1993rv}
M.~Maggiore, ``{A Generalized uncertainty principle in quantum gravity},''
  \href{http://dx.doi.org/10.1016/0370-2693(93)91401-8}{{\em Phys. Lett. B}
  {\bfseries 304} (1993) 65--69},
  \href{http://arxiv.org/abs/hep-th/9301067}{{\ttfamily arXiv:hep-th/9301067}}.

\bibitem{Adler:1999bu}
R.~J. Adler and D.~I. Santiago, ``{On gravity and the uncertainty principle},''
  \href{http://dx.doi.org/10.1142/S0217732399001462}{{\em Mod. Phys. Lett. A}
  {\bfseries 14} (1999) 1371},
  \href{http://arxiv.org/abs/gr-qc/9904026}{{\ttfamily arXiv:gr-qc/9904026}}.

\bibitem{Hossenfelder:2012jw}
S.~Hossenfelder, ``{Minimal Length Scale Scenarios for Quantum Gravity},''
  \href{http://dx.doi.org/10.12942/lrr-2013-2}{{\em Living Rev. Rel.}
  {\bfseries 16} (2013) 2}, \href{http://arxiv.org/abs/1203.6191}{{\ttfamily
  arXiv:1203.6191 [gr-qc]}}.

\bibitem{Hagar:2014}
A.~Hagar, {\em {Discrete or Continuous? The Quest for Fundamental Length in
  Modern Physics}}.
\newblock Cambridge University Press, 2014.

\bibitem{Maggiore:1993kv}
M.~Maggiore, ``{The Algebraic structure of the generalized uncertainty
  principle},'' \href{http://dx.doi.org/10.1016/0370-2693(93)90785-G}{{\em
  Phys. Lett. B} {\bfseries 319} (1993) 83--86},
  \href{http://arxiv.org/abs/hep-th/9309034}{{\ttfamily arXiv:hep-th/9309034}}.

\bibitem{Maggiore:1993zu}
M.~Maggiore, ``{Quantum groups, gravity and the generalized uncertainty
  principle},'' \href{http://dx.doi.org/10.1103/PhysRevD.49.5182}{{\em Phys.
  Rev. D} {\bfseries 49} (1994) 5182--5187},
  \href{http://arxiv.org/abs/hep-th/9305163}{{\ttfamily arXiv:hep-th/9305163}}.

\bibitem{Kempf:1994su}
A.~Kempf, G.~Mangano, and R.~B. Mann, ``{Hilbert space representation of the
  minimal length uncertainty relation},''
  \href{http://dx.doi.org/10.1103/PhysRevD.52.1108}{{\em Phys. Rev. D}
  {\bfseries 52} (1995) 1108--1118},
  \href{http://arxiv.org/abs/hep-th/9412167}{{\ttfamily arXiv:hep-th/9412167}}.

\bibitem{Maggiore:2002qr}
M.~Maggiore, ``{The Atick-Witten free energy, closed tachyon condensation and
  deformed Poincar\'e symmetry},''
  \href{http://dx.doi.org/10.1016/S0550-3213(02)00938-0}{{\em Nucl. Phys. B}
  {\bfseries 647} (2002) 69--100},
  \href{http://arxiv.org/abs/hep-th/0205014}{{\ttfamily arXiv:hep-th/0205014}}.

\bibitem{Amelino-Camelia:2013fxa}
G.~Amelino-Camelia, ``{Challenge to Macroscopic Probes of Quantum Spacetime
  Based on Noncommutative Geometry},''
  \href{http://dx.doi.org/10.1103/PhysRevLett.111.101301}{{\em Phys. Rev.
  Lett.} {\bfseries 111} (2013) 101301},
  \href{http://arxiv.org/abs/1304.7271}{{\ttfamily arXiv:1304.7271 [gr-qc]}}.

\bibitem{Nouicer:2007jg}
K.~Nouicer, ``{Quantum-corrected black hole thermodynamics to all orders in the
  Planck length},''
  \href{http://dx.doi.org/10.1016/j.physletb.2006.12.072}{{\em Phys. Lett. B}
  {\bfseries 646} (2007) 63--71},
  \href{http://arxiv.org/abs/0704.1261}{{\ttfamily arXiv:0704.1261 [gr-qc]}}.

\bibitem{Pedram:2011gw}
P.~Pedram, ``{A Higher Order GUP with Minimal Length Uncertainty and Maximal
  Momentum},'' \href{http://dx.doi.org/10.1016/j.physletb.2012.07.005}{{\em
  Phys. Lett. B} {\bfseries 714} (2012) 317--323},
  \href{http://arxiv.org/abs/1110.2999}{{\ttfamily arXiv:1110.2999 [hep-th]}}.

\bibitem{Chung:2019raj}
W.~S. Chung and H.~Hassanabadi, ``{A new higher order GUP: one dimensional
  quantum system},''
  \href{http://dx.doi.org/10.1140/epjc/s10052-019-6718-3}{{\em Eur. Phys. J. C}
  {\bfseries 79} no.~3, (2019) 213}.

\bibitem{Petruzziello:2020een}
L.~Petruzziello, ``{Generalized uncertainty principle with maximal observable
  momentum and no minimal length indeterminacy},''
  \href{http://dx.doi.org/10.1088/1361-6382/abfd8f}{{\em Class. Quant. Grav.}
  {\bfseries 38} no.~13, (2021) 135005},
  \href{http://arxiv.org/abs/2010.05896}{{\ttfamily arXiv:2010.05896
  [hep-th]}}.

\bibitem{Magueijo:2002am}
J.~Magueijo and L.~Smolin, ``{Generalized Lorentz invariance with an invariant
  energy scale},'' \href{http://dx.doi.org/10.1103/PhysRevD.67.044017}{{\em
  Phys. Rev. D} {\bfseries 67} (2003) 044017},
  \href{http://arxiv.org/abs/gr-qc/0207085}{{\ttfamily arXiv:gr-qc/0207085}}.

\bibitem{Jizba:2009qf}
P.~Jizba, H.~Kleinert, and F.~Scardigli, ``{Uncertainty Relation on World
  Crystal and its Applications to Micro Black Holes},''
  \href{http://dx.doi.org/10.1103/PhysRevD.81.084030}{{\em Phys. Rev. D}
  {\bfseries 81} (2010) 084030},
  \href{http://arxiv.org/abs/0912.2253}{{\ttfamily arXiv:0912.2253 [hep-th]}}.

\bibitem{Bonechi:1992qf}
F.~Bonechi, E.~Celeghini, R.~Giachetti, E.~Sorace, and M.~Tarlini,
  ``{Inhomogeneous quantum groups as symmetries of phonons},''
  \href{http://dx.doi.org/10.1103/PhysRevLett.68.3718}{{\em Phys. Rev. Lett.}
  {\bfseries 68} (1992) 3718--3720},
  \href{http://arxiv.org/abs/hep-th/9201002}{{\ttfamily arXiv:hep-th/9201002}}.

\bibitem{Lukierski:1991}
J.~Lukierski, H.~Ruegg, A.~Nowicki, and V.~Tolstoy, ``{q-deformation of
  Poincar\'e algebra},'' {\em Phys. Lett. B} {\bfseries 264} (1991) 331.

\bibitem{Lukierski:1992dt}
J.~Lukierski, A.~Nowicki, and H.~Ruegg, ``{New quantum Poincare algebra and k
  deformed field theory},''
  \href{http://dx.doi.org/10.1016/0370-2693(92)90894-A}{{\em Phys. Lett. B}
  {\bfseries 293} (1992) 344--352}.

\bibitem{Lukierski:1993wxa}
J.~Lukierski and H.~Ruegg, ``{Quantum kappa Poincare in any dimension},''
  \href{http://dx.doi.org/10.1016/0370-2693(94)90759-5}{{\em Phys. Lett. B}
  {\bfseries 329} (1994) 189--194},
  \href{http://arxiv.org/abs/hep-th/9310117}{{\ttfamily arXiv:hep-th/9310117}}.

\bibitem{Pikovski:2011zk}
I.~Pikovski, M.~R. Vanner, M.~Aspelmeyer, M.~S. Kim, and C.~Brukner, ``{Probing
  Planck-scale physics with quantum optics},''
  \href{http://dx.doi.org/10.1038/nphys2262}{{\em Nature Phys.} {\bfseries 8}
  (2012) 393--397}, \href{http://arxiv.org/abs/1111.1979}{{\ttfamily
  arXiv:1111.1979 [quant-ph]}}.

\bibitem{Kumar:2019bnd}
S.~P. Kumar and M.~B. Plenio, ``{On Quantum Gravity Tests with Composite
  Particles},'' \href{http://dx.doi.org/10.1038/s41467-020-17518-5}{{\em Nature
  Commun.} {\bfseries 11} no.~1, (2020) 3900},
  \href{http://arxiv.org/abs/1908.11164}{{\ttfamily arXiv:1908.11164
  [quant-ph]}}.

\bibitem{Pedram:2011aa}
P.~Pedram, ``{New Approach to Nonperturbative Quantum Mechanics with Minimal
  Length Uncertainty},''
  \href{http://dx.doi.org/10.1103/PhysRevD.85.024016}{{\em Phys. Rev. D}
  {\bfseries 85} (2012) 024016},
  \href{http://arxiv.org/abs/1112.2327}{{\ttfamily arXiv:1112.2327 [hep-th]}}.

\bibitem{Ali:2009zq}
A.~F. Ali, S.~Das, and E.~C. Vagenas, ``{Discreteness of Space from the
  Generalized Uncertainty Principle},''
  \href{http://dx.doi.org/10.1016/j.physletb.2009.06.061}{{\em Phys. Lett. B}
  {\bfseries 678} (2009) 497--499},
  \href{http://arxiv.org/abs/0906.5396}{{\ttfamily arXiv:0906.5396 [hep-th]}}.

\bibitem{Ali:2011fa}
A.~F. Ali, S.~Das, and E.~C. Vagenas, ``{A proposal for testing Quantum Gravity
  in the lab},'' \href{http://dx.doi.org/10.1103/PhysRevD.84.044013}{{\em Phys.
  Rev. D} {\bfseries 84} (2011) 044013},
  \href{http://arxiv.org/abs/1107.3164}{{\ttfamily arXiv:1107.3164 [hep-th]}}.

\end{thebibliography}\endgroup

\end{document}